# Nearly twofold overestimation of the superconducting volume fraction in pressurized Ruddlesden-Popper nickelates

Aleksandr V. Korolev[*] and Evgeny F. Talantsev[**]

M.N. Miheev Institute of Metal Physics, Ural Branch, Russian Academy of Sciences, 18, S. Kovalevskaya St., Ekaterinburg, 620108, Russia

* korolyov@imp.uran.ru
** evgeny.talantsev@imp.uran.ru

**Abstract**

The detection of the DC diamagnetic state in pressurized Ruddlesden-Popper nickelates remained an unsolved experimental problem until recent experiments in which Zhu et al.[1] measured the DC diamagnetic responses in zero-field cooled (ZFC) mode in pressurized $La_4Ni_3O_{10}$. Zhu et al.[1] reported that the ratio of the measured ZFC magnetic moment to the Meissner magnetic moment of the sample (and this ratio was termed the superconducting volume fraction $f$) reaches 81–86%. We regard outstanding experimental results[1]; however, our calculations based on the reported experimental datasets[1] using the standard procedure showed the ratio to be 51–59%. Upon our request, Zhu et al.[1] provided detailed explanations and the equation they used to calculate $f$. To our knowledge, the procedure[1] and equation[1] for calculating $f$ have never been mentioned or described before, including in Ref.[1]. Here we argue that the proposed equation[1] and procedure[1] are incorrect, and that this equation[1] results in multiple overestimations of the superconducting volume fraction in the sample. This overestimation error affects all superconducting volume fractions $f$ in Ruddlesden-Popper nickelates reported to date[1–4]. Therefore, we describe the error we discovered in this paper.

# Nearly twofold overestimation of the superconducting volume fraction in pressurized Ruddlesden-Popper nickelates

First of all, we need to note that Zhu et al.[1] used Gaussian units for ZFC magnetization data and SI units for magnetoresistance data. Sometimes, the mixture of Gaussian and SI systems causes tenfold mistakes[5]. In addition, Zhu et al.[1] used the letter $M$ to denote the magnetic moment of the sample, which is incorrect in both the Gaussian and SI systems[6], because $M$ is reserved for magnetization. In this paper, we have converted the ZFC magnetization data[1] for Sample S6 to SI units. Zhu et al.[1] also presented ZFC data and plots for samples S5 and S7, but the dimensions and shape of these samples were not reported, and thus we could not able to perform data analysis for these samples.

In Figure 1 we show the Extended Data Figure S9(a-d)[1] for the $La_4Ni_3O_{10}$ (sample S6[1]), where it is shown that the ratio of the $m_{ZFC,measured}$ (which is experimentally measured magnetic moment of the sample in ZFC mode in the applied field $H$,) to the $m_{Meissner}$ (which is the calculated magnetic moment of the sample in the assumption that the sample (having its physical sizes, shape, and volume) in the applied field $H$ is in the Meissner state) is:

$$0.81 \leq \frac{|m_{ZFC,measured}|}{|m_{Meissner}|} \leq 0.86. \qquad (1)$$

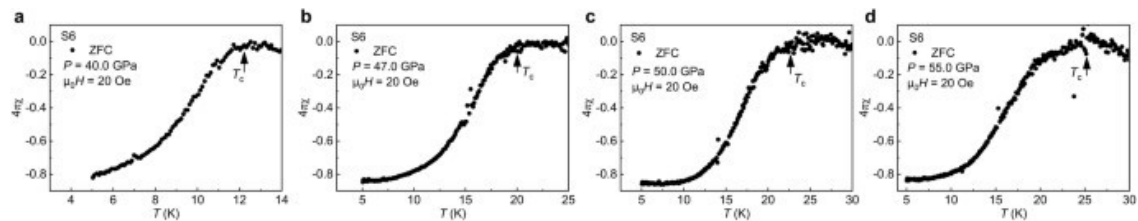

**FIGURE 1**. ZFC magnetization data reported by Zhu et al.[1] for Sample S6 in their Extended Data Figure S9(a-d)[1], where the ratio of the measured magnetic moment of the sample to the Meissner magnetic moment is higher than 80% for all panels. Reproduced with permission from Springer Nature Licence office (Licence number: 6214151315070, February 22, 2026).

Figure 2 shows our calculations, which are based on the standard procedure[7–14] (described in Supplementary Materials) and for which we used raw experimental data $m_{ZFC,measured}$ reported for sample S6[1] in Figure 3(f-i)[1]. As can be seen (Figure 2), our calculations show:

$$0.51 \leq \frac{|m_{ZFC,measured}|}{|m_{Meissner}|} \leq 0.59. \quad (2)$$

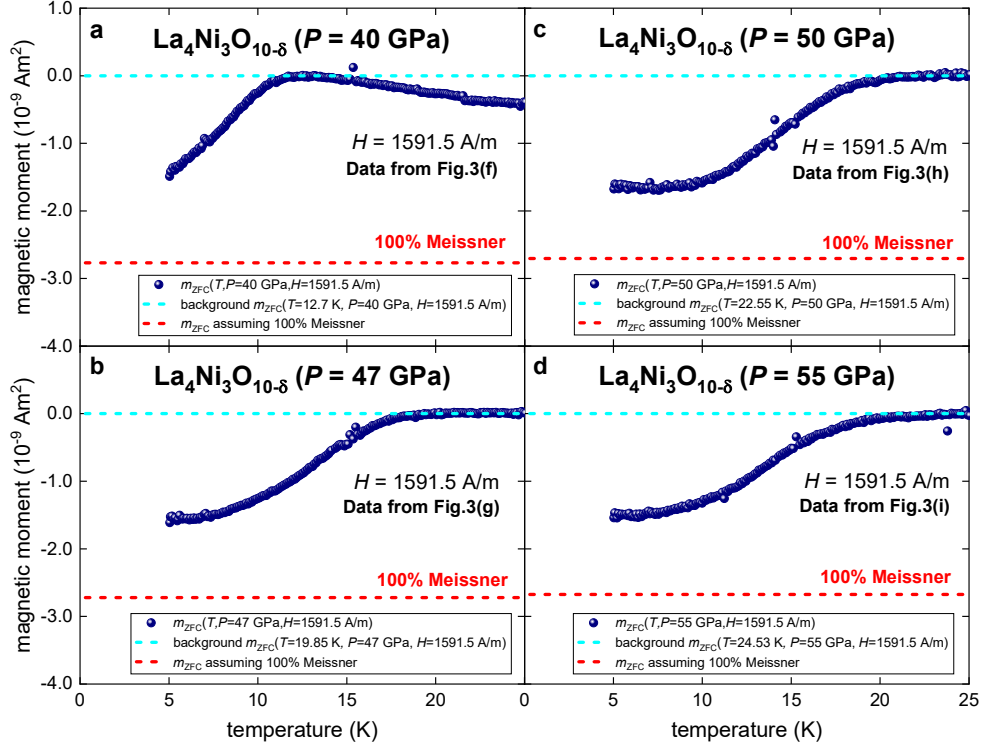

**FIGURE 2.** $m_{ZFC,measured}\left(P, H = 1591.5 \frac{A}{m}, T\right)$ curves in La4Ni3O10-δ sample S6[1] (navy) and calculated $m_{Meissner}$ for sample S6 (disk sample with a diameter $d$ = 160 μm and a thickness $h$ = 22 μm) at given pressure and applied field $H = 1591.5 \frac{A}{m}$ (red). (**a**) $P = 40\ GPa$, (**b**) $P = 47\ GPa$, (**c**) $P = 50\ GPa$, (**d**) $P = 55\ GPa$.

In friendly response to our request to explain the disagreement, the authors[1] informed us that in all their figures[1] and reported values[1] they calculated the volume fraction of the superconducting phase using the following equation:

$$f = \frac{\frac{|m_{ZFC,measured}|}{|m_{Meissner}|}}{1 + N \times \left(\frac{|m_{ZFC,measured}|}{|m_{Meissner}|} - 1\right)}, \quad (3)$$

where $N$ is the magnetometric[7] demagnetization factor[10–13] of the sample calculated from the physical sizes of the sample and its shape. The derivation of Equation 3 can be provided by the authors[1].

Here we need to clarify that, to the best of our knowledge:

1. Equation 3[1] has never been used prior to Ref.[1];
2. The derivation of this equation was not presented in Ref.[1];
3. The use of this equation was not mentioned in Ref.[1].

Thus, our task was to test the validity of Equation 3. In the result of our examination, we report that Equation 3 is incorrect, and the calculated values $f$ are not related to the volume fraction of the superconducting phase in the $La_4Ni_3O_{10}$[1] samples.

We prove our statement by the following consideration. Zhu et al.[1] reported that $La_4Ni_3O_{10}$ single crystal (sample S6[1]) had a disk shape with a diameter $d = 160$ μm and a thickness $h = 22$ μm. For a superconductor in the Meissner state[15], the magnetic flux density inside the sample is zero by definition, $B = 0$. Thus, the primary equation of the magnetostatic is:

$$B = 0 = \mu_0 \times (M_V + H - N \times M_V), \quad (4)$$

$$M_V = -\frac{H}{1-N}, \quad (5)$$

where $\mu_0$ is the magnetic permeability of free-space (in units of $[NA^{-2}]$), $H$ is the applied magnetic field (in units of $\left[\frac{A}{m}\right]$), and $M_V$ is the sample volume magnetization (in units of $\left[\frac{A}{m}\right]$), which is calculated from the magnetic moment of the sample in the Meissner state $m_{Meissner}$ (in units of $[Am^2]$), and the volume of the sample $V$ (in units of $[m^3]$):

$$M_V = \frac{m_{Meissner}}{V}. \quad (6)$$

Combining Eqs. 5,6, the measured in the experiment magnetic moment of the sample $m$ in the Meissner state is:

$$m_{Meissner} = -V \times \frac{H}{1-N}. \tag{7}$$

Eq. 7 is the fundamental equation in superconductivity, which is true for all superconductors (see, for instance, Ref.[12,14]). We acknowledged a fact that Zhu et al.[1] agreed with Equations 4-7.

The demagnetization factor, $N$, for a disk shape sample can be accurately calculated by the equation[10]:

$$N = 1 - \frac{1}{1+q \times \frac{d}{h}}, \tag{8}$$

$$q = \frac{4}{3\pi} + \frac{2}{3\pi} \times tanh\left(1.27 \times \frac{h}{d} \times ln\left(1 + \frac{d}{h}\right)\right) \tag{9}$$

The substituting sizes for sample S6[1] in Eqs. 8,9 yields:

$$N = 0.784 \tag{10}$$

Zhu et al.[1] used a simpler approximation[13] to calculate $N^{10}$. The approach[13] was also used in Refs.[3,4]; however, this is not a source for our disagreement, and this was confirmed by the authors[1,3].

Zhu et al.[1] used the applied field $H = 1591.5 \frac{A}{m}$ in their ZFC experiments, and thus sample S6[1] has, in an assumption of 100% Meissner magnetic moment:

$$m_{Meissner} = -V \times \frac{H}{1-N} = -\left(\frac{\pi}{4} \times (1.6 \times 10^{-4})^2 \times (2.2 \times 10^{-5}) \, m^3\right) \times \frac{1591.5 \frac{A}{m}}{1-0.784} =$$

$$-3.26 \times 10^{-9} \, Am^2 \tag{11}$$

The correctness of Eq. 11 was confirmed by the authors[1].

For the analysis of Equation 3, let us consider a case when Sample S6[1] contains 50% superconducting phase by volume. Two distributions of the superconducting phase in Sample A and Sample B are shown in Figure 3.

Sample A has a superconducting part as a disk with $d = 160 \, \mu m$ and $h = 11 \, \mu m$, and Sample B has a superconducting part as a disk with $d = 113.2 \, \mu m$ and $h = 22 \, \mu m$.

It should be emphasized that since the superconducting phase content is 50% in both Samples A and B, Equation 3 should give $f = 0.50$ for Sample A and for Sample B.

Let us make calculations for Sample A. Measured ZFC magnetic moment of Sample A in the field $H = 1591.5 \frac{A}{m}$ is:

$$m_{ZFC, measured\ Sample\ A} = -V_{supercond.\ part\ of\ Sample\ A} \times \frac{H}{1-N_{Sample\ A}} = \left(\frac{\pi}{4} \times \right.$$

$$\left. (1.6 \times 10^{-4})^2 \times (1.1 \times 10^{-5})\ m^3 \right) \times \frac{1591.5 \frac{A}{m}}{1-0.8734} = -2.78 \times 10^{-9}\ Am^2. \qquad (12)$$

where $N = 0.8734$ is calculated by substituting $d = 160\ \mu m$ and $h = 11\ \mu m$ in Equations 8.9. In the result,

$$\frac{|m_{ZFC,measured}|}{|m_{Meissner}|} = \frac{2.78 \times 10^{-9}\ Am^2}{3.26 \times 10^{-9}\ Am^2} = 0.853 \qquad (13)$$

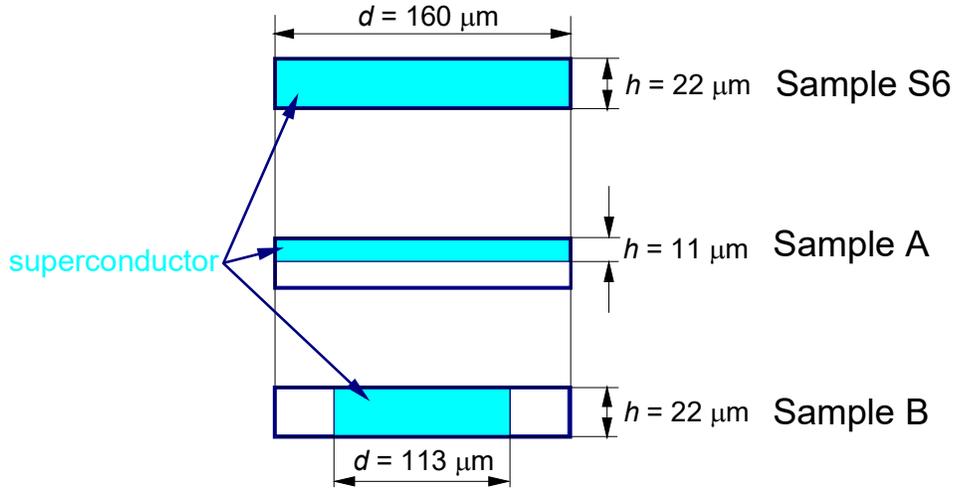

**FIGURE 3**. Cross-sections of superconducting disks with the same outer diameters $d = 160\ \mu m$ and thicknesses $h = 22\ \mu m$ as Sample S6[1]. Top: the superconductor fills the entire disk (this sample is designated in the text as Sample S6). Middle: the superconductor fills 50% of the disk in the form of a disk with a diameter $d = 160\ \mu m$ and thickness $h = 11\ \mu m$ (this sample is designated in the text as Sample A). Bottom: the superconductor fills 50% of the disk in the form of a disk with a diameter $d = 113\mu m$ and thickness $h = 22\ \mu m$ (this sample is designated in the text as Sample B).

Substituting the values from Equations 10 and 13 into Equation 3 we obtain:

$$f = \frac{\frac{|m_{ZFC,measured,Sample\ A}|}{|m_{Meissner}|}}{1 + N \times \left(\frac{|m_{ZFC,measured,Sample\ A}|}{|m_{Meissner}|} - 1\right)} = \frac{0.853}{1 + 0.784 \times (0.853 - 1)} = 0.964. \qquad (14)$$

Thus, Equation 3[1] fails to establish that sample A has a 50% volume fraction filled by superconducting phase. In contrast, the calculation by Equation 3[1] overestimates the superconducting phase volume fraction by a factor of ~2, and the calculated fraction is almost equal to 100%.

Let us make calculations for Sample B. Measured ZFC magnetic moment of Sample B in the field $H = 1591.5 \frac{A}{m}$ is:

$$m_{ZFC, measured\ Sample\ B} = -V_{supercond.\ part\ of\ Sample\ B} \times \frac{H}{1-N_{Sample\ A}} = \left(\frac{\pi}{4} \times (1.132 \times 10^{-4})^2 \times (2.2 \times 10^{-5})\ m^3\right) \times \frac{1591.5 \frac{A}{m}}{1-0.7255} = -1.284 \times 10^{-9}\ Am^2. \quad (15)$$

where $N = 0.7255$ is calculated by substituting $d = 113.2\ \mu m$ and $h = 22\ \mu m$ in Equations 8,9. In the result,

$$\frac{|m_{ZFC, measured, Sample\ B}|}{|m_{Meissner}|} = \frac{1.284 \times 10^{-9}\ Am^2}{3.26 \times 10^{-9}\ Am^2} = 0.394 \quad (16)$$

Substituting the values from Equations 10 and 16 into Equation 3 we obtain:

$$f = \frac{\frac{|m_{ZFC, measured, Sample\ B}|}{|m_{Meissner}|}}{1+N \times \left(\frac{|m_{ZFC, measured, Sample\ B}|}{|m_{Meissner}|}-1\right)} = \frac{0.394}{1+0.784 \times (0.394 - 1)} = 0.751. \quad (17)$$

Thus, equation 3[1] fails to yield 50% superconducting phase fraction in Sample B. Instead, it overestimates the superconducting phase fraction in Sample B by 1.5 times.

In summary, we have shown that method[1] and equation[1] for calculating the superconducting volume fraction *f* in Ruddlesden-Popper nickelate samples, used *de facto* (and without the derivation of the equation) in Reference[1] and also in References[2–4] (also without derivation), are incorrect. This approach[1] significantly overestimates the superconducting volume fractions *f* in all Ruddlesden-Popper nickelate samples reported to date[1–4].

**Author contributions**

AVK and EFT jointly conceived the work; AVK analysed data and performed the calculations; EFT confirmed the calculations and wrote the manuscript, which was revised by AVK.

**Acknowledgements**

The work was carried out within the framework of the state assignment of the Ministry of Science and Higher Education of the Russian Federation for the IMP UB RAS.

**Competing interests**

The authors declare no competing interests.

*Supplementary Materials*

**Nearly twofold overestimation of the superconducting volume fraction in pressurized Ruddlesden-Popper nickelates**

Aleksandr V. Korolev[*] and Evgeny F. Talantsev[**]

M.N. Miheev Institute of Metal Physics, Ural Branch, Russian Academy of Sciences,
18, S. Kovalevskoy St., Ekaterinburg, 620108, Russia

* korolyov@imp.uran.ru
** evgeny.talantsev@imp.uran.ru

Zhu et al.[1] used Gaussian units for ZFC magnetization data. In addition, Zhu et al.[1] used the letter $M$ to denote the magnetic moment of the sample, which is incorrect in both the Gaussian and SI systems[6], because $M$ is reserved for magnetization. Here we have converted the ZFC magnetization data[1] to SI units.

Zhu et al.[1] reported that $La_4Ni_3O_{10}$ single crystal (sample S6) had a disk shape with a diameter $d = 160$ μm and a thickness $h = 22$ μm.

For a superconductor in the Meissner state[15], the magnetic flux density inside the sample is zero by definition, $B = 0$, and for a sample with magnetometric[7] demagnetization factor[10–13] $N$, the primary equation of the magnetostatic is:

$$B = 0 = \mu_0 \times (M_V + H - N \times M_V), \qquad (A1)$$

$$M_V = -\frac{H}{1-N}, \qquad (A2)$$

where $\mu_0$ is the magnetic permeability of free-space (in units of $[NA^{-2}]$), $H$ is the applied magnetic field (in units of $\left[\frac{A}{m}\right]$), and $M_V$ is the sample volume magnetization (in units of $\left[\frac{A}{m}\right]$), which is calculated from the measured sample magnetic moment $m$ (in units of $[Am^2]$), and the volume of the sample $V$ (in units of $[m^3]$):

$$M_V = \frac{m}{V}. \qquad (A3)$$

Combining Eqs. A1-A3, the measured magnetic moment of the sample $m$ in the Meissner state is:

$$m_{Meissner} = -V \times \frac{H}{1-N}. \tag{A4}$$

Eq. A4 is the fundamental equation in superconductivity, which is in use for all superconductors (see, for instance, Ref.[12,14]).

The demagnetization factor, $N$, for a disk shape sample can be accurately calculated by the equation[10]:

$$N = 1 - \frac{1}{1+q \times \frac{d}{h}}, \tag{A5}$$

$$q = \frac{4}{3\pi} + \frac{2}{3\pi} \times \tanh\left(1.27 \times \frac{h}{d} \times \ln\left(1 + \frac{d}{h}\right)\right) \tag{A6}$$

The substituting sizes for sample S6 in Eqs. A5, A6 yields $N = 0.784$.

We agree with Zhu et al.[1] that as the first order of approximation, one can assume that the sample exhibits the proportional contraction of the lattice dimensions under the pressure, and, based on that, the demagnetization factor remains unchanged under hydrostatic pressure conditions.

Because Equation A4 has the sample volume, which depends on pressure, the final equation for the magnetic moment of the pressurized sample in the Meissner state is:

$$m_{Meissner}(P, H) = -V_s(P) \times \frac{H}{1-N} = -V_0 \times \frac{V_{u.c.}(P)}{V_{0,u.c.}} \times \frac{H}{1-N} \tag{A7}$$

where $V_0$ is sample volume at ambient pressure, $V_{0,u.c.}$ is unit cell volume at ambient conditions, $V_{u.c.}(P)$ is the unit cell volume at pressure $P$.

Zhu et al.[1] reported the $V_{u.c.}(P)$ dataset and its approximation by Birch-Murnaghan model[16,17] for $La_4Ni_3O_{10}$ single crystal.

Based on known values, the magnetic moment of the single crystal La$_4$Ni$_3$O$_{10}$ (S6) in the Meissner state in the applied field $H = 1591.5 \frac{A}{m}$ and pressure $P = 40\ GPa$ (these conditions were applied for sample S6 in Fig. 3,f [1]) is:

$$m_{Meissner}\left(P = 40\ GPa, H = 1591.5\ \frac{A}{m}\right)[Am^2] = -V[m^3] \times \frac{H\left[\frac{A}{m}\right]}{1-N} = -\left(\frac{\pi}{4} \times \right.$$

$$\left. (1.6 \times 10^{-4})^2 \times (2.2 \times 10^{-5})\ m^3\right) \times \frac{(3.51 \times 10^{-28})\ m^3}{(4.136 \times 10^{-28})\ m^3} \times \frac{1591.5\ \frac{A}{m}}{1-0.784} = -2.77 \times 10^{-9}\ Am^2.$$

(A8)

Since a diamond anvil cell (DAC) acquires its own magnetic moment in an applied magnetic field $H$, to determine the sample magnetic moment $m$, the background magnetic moment generated by the DAC must be subtracted from the measured magnetic moment. Here we used the approach used by Zhu et al.[1], which is to assume that the DAC magnetic moment is temperature independent at temperatures lower than the onset of the superconducting transition. In Fig. 1 (main text) we performed all the described above calculations for Sample S6[1] (for which measured data are shown in Fig. 3,f-i[1]). For instance:

$$m_{ZFC,measured}\left(P = 40\ GPa, H = 1591.5\ \frac{A}{m}, T = 5.098\ K\right) = -1.41 \times 10^{-9}\ Am^2,$$

(A9)

and the ratio:

$$\frac{m_{ZFC,measured}\left(P=40\ GPa, H=1591.5\frac{A}{m}, T=5.098\ K\right)}{m_{Meissner}\left(P=40\ GPa, H=1591.5\frac{A}{m}\right)} = \frac{1.41 \times 10^{-9}\ Am^2}{2.77 \times 10^{-9}\ Am^2} = 0.509 \cong 51\%$$

(A10)

Despite Zhu et al.[1] not providing data for all panels in their Extended Data Fig. 9[1], the figure in Extended Data Fig. 9(a)[1] shows that at pressure $P = 40\ GPa$ the ratio is $\cong 81\%$. This large difference between calculations by Zhu et al.[1] and our Equation A10 is the demonstration of our disagreement with Zhu et al.[1]

All $m_{ZFC}\left(P, H = 1591.5 \frac{A}{m}, T\right)$ dataset for sample S6, which we recalculated from raw data[1] by the described method, are shown in Figure 1 in main text, where each panel shows the $m_{Meissner}\left(P, H = 1591.5 \frac{A}{m}\right)$ values. One can see that the measured diamagnetic moment at all pressures is:

$$50.9\% \leq \frac{m_{ZFC,measured}(P,H,T)}{m_{Meissner}(P,H)} \leq 59.2\% \quad (A11)$$

Now we need to describe our fundamental disagreement with Zhu et al.[1] and other teams[3,4] who used ratios similar to Equations A10, A11 as the volume fraction of the superconducting phase in the sample.

Equations A1-A7 were derived under the assumption that the superconducting phase is solid, continues, exhibits 100% diamagnetism and fills the entire sample volume. For these conditions, the magnetometric[7] demagnetization factor[10–13] $N$ was calculated. This means that Equations 10,11 can only be used for the confirmation that the measured sample (which exhibits magnetic moment $m_{ZFC,measured}$) has 100% superconducting phase:

$$\frac{m_{ZFC,measured}(P,H,T)}{m_{s,Meissner}(P,H)} = 100\%. \quad (A12)$$

Several percent deviation from 100% is an acceptable level[14] of experimental inaccuracy.

However, if the ratio is:

$$\frac{m_{ZFC,measured}(P,H,T)}{m_{Meissner}(P,H)} = 59.2\%, \quad (A13)$$

then there is a principal uncertainty in the superconducting volume fraction in the sample.

Let us demonstrate this with one example. Sample S6 at $P = 50\ GPa$ exhibit magnetic moment (data is in Fig. 1,c (in our main text) and Fig. 3,h[1]):

$$m_{ZFC,measured}\left(P = 50\ GPa, H = 1591.5 \frac{A}{m}, T = 5.098\ K\right) = -1.60 \times 10^{-9}\ Am^2$$

$$(A14)$$

If one assumes that this sample S6 contained only a superconducting lamella with disk shape with diameter $d = 145$ μm and thickness $h = 6.5$ μm. The volume of this lamella is 24.3% of the sample S6 volume (because sample S6 has diameter $d = 160$ μm and thickness $h = 22$ μm). However, this lamella in the Meissner state and the applied field $H = 1591.5 \frac{A}{m}$ and pressure $P = 50\ GPa$ has the same magnetic moment as the magnetic moment measured by Zhu et al.[1] (for which the authors[1] claimed that the sample has 86% of superconducting fraction):

$$m_{ZFC,measured}\left(P = 50\ GPa, H = 1591.5\ \tfrac{A}{m}\right)[Am^2] = -V_s[m^3] \times \frac{H\left[\tfrac{A}{m}\right]}{1-N} =$$

$$-\left(\tfrac{\pi}{4} \times (1.45 \times 10^{-4})^2 \times (6.5 \times 10^{-6})\ m^3\right) \times \frac{(3.428 \times 10^{-28})\ m^3}{(4.136 \times 10^{-28})\ m^3} \times \frac{1591.5\ \tfrac{A}{m}}{1-0.9116} = -1.60 \times 10^{-9}\ Am^2 \tag{A15}$$

where $N = 0.9116$ was calculated by Equations A5, A6.

We need to note that there are an infinite number of lamella sizes, smaller than the physical sizes of sample S6, that exhibit the same magnetic moment as the measured in experiment (Equation A14).